\documentclass[10pt,english,aps,preprint,prb,floatfix,superscriptaddress,twocolumn,showpacs,amsmath,amssymb,reprint]{revtex4-1}
\usepackage[T1]{fontenc}
\usepackage[latin9]{inputenc}
\usepackage{color}
\usepackage{amsbsy}
\usepackage{amssymb}
\usepackage{graphicx}
\usepackage{esint}
\PassOptionsToPackage{normalem}{ulem}
\usepackage{ulem}
\makeatletter

\@ifundefined{textcolor}{}
{
 \definecolor{BLACK}{gray}{0}
 \definecolor{WHITE}{gray}{1}
 \definecolor{RED}{rgb}{1,0,0}
 \definecolor{GREEN}{rgb}{0,1,0}
 \definecolor{BLUE}{rgb}{0,0,1}
 \definecolor{CYAN}{cmyk}{1,0,0,0}
 \definecolor{MAGENTA}{cmyk}{0,1,0,0}
 \definecolor{YELLOW}{cmyk}{0,0,1,0}
}

\PassOptionsToPackage{caption=false}{subfig}

\makeatother
\usepackage{amsmath}

\makeatother

\usepackage{babel}
\begin{document}

\title{
Effective response theory for zero energy Majorana bound states in three spatial dimensions
}

\author{Pedro L. e S. Lopes}

\email{plslopes@ifi.unicamp.br}

\affiliation{Instituto de F\'{i}sica Gleb Wataghin, Universidade Estadual de Campinas,
Campinas, SP 13083-970, Brazil}

\affiliation{Department of Physics and Institute for Condensed Matter Theory,
University of Illinois, 1110 W. Green St., Urbana IL 61801-3080, U.S.A.}

\author{Jeffrey C. Y. Teo}

\affiliation{Department of Physics, University of Virginia, Charlottesville, VA 22904, U.S.A.}

\author{Shinsei Ryu}

\affiliation{Department of Physics and Institute for Condensed Matter Theory,
University of Illinois, 1110 W. Green St., Urbana IL 61801-3080, U.S.A.}

\begin{abstract}

We propose a gravitational response theory for point defects (hedgehogs) binding Majorana zero modes in 
(3+1)-dimensional superconductors. Starting in 4+1 dimensions, where the point defect is extended into a line, a coupling of the bulk defect texture with 
the gravitational  field is introduced. Diffeomorphism invariance then leads to an $SU(2)_2$ Kac-Moody current running along the 
defect line. The $SU(2)_2$ Kac-Moody algebra accounts for the non-Abelian nature of the zero modes  in 3+1 dimensions.
It is then shown to also encode the angular momentum density which permeates throughout the bulk between hedgehog-anti-hedgehog pairs.
\end{abstract}

\maketitle

\section{Introduction}

Topological phases are gapped quantum phases of matter,   
which are impossible to be characterized via spontaneous symmetry breaking.
The list of known topological states in condensed matter has been greatly expanded recently;
in addition to quantum Hall states which feature a genuine, intrinsic topological order
which do not require any symmetry, 
topological states with (or protected by) symmetries,
such as symmetry-protected and symmetry-enriched topological phases, 
have been discovered and extensively discussed recently.
\cite{HassanKane2010, QiZhang2011, PhysRevB.87.155114}
For non-interacting fermion systems with a certain set of discrete symmetries, 
the classification of all topological phases is possible and 
summarized in terms of a periodic table. 
\cite{Schnyder2008, NewJPhysRyu,Kitaevtable08}

Topological defects introduce further complexity and possibilities in topological phases of matter.
\cite{PhysRevB.82.115120} 
For instance, while the periodic table does not list topological superconductors with broken time-reversal symmetry in three spatial dimensions,
\cite{NewJPhysRyu,Kitaevtable08} one can endow trivial superconductors with non-trivial topological properties 
by introducing point defects. 
These defects may be realized, for example, as superconducting vortices on the surface of topological insulators with proximity-induced superconductivity. 
\cite{FuKane08} These topological defects host zero energy Majorana bound states (MBS) at their cores which are robust against any perturbation weaker than the bulk energy gap,
and are shown to obey non-Abelian statistics in (3+1) dimensions. 
\cite{PhysRevLett.104.046401, PhysRevB.83.115132, PhysRevB.84.245119} 

While Ref.\ \onlinecite{PhysRevLett.104.046401}
provides the descriptions of such topological defects in terms of single-particle Hamiltonians, 
characterization of the defects beyond the non-interacting limit remains a challenge. 
Often, a good way to tackle this problem comes from appealing to topological field theories.
\cite{QiHughesZhang08,PhysRevB.85.045104} 
They are desirable as they directly suggest the presence of boundary excitations for topological insulators
\cite{QiHughesZhang08} and the quasiparticle braiding behavior in (2+1)D topologically ordered phases
\cite{WenZee92}. They are usually related to bulk-boundary anomalies, surviving the effects of interactions and giving the phenomenological 
responses expected from the low-energy excitations.

It is our objective here to propose and analyze a topological field
theory that describes point defects in (3+1)D superconductors with broken time-reversal symmetry, 
belonging to the symmetry class D in Altland-Zirnbauer classification.
\cite{AltlandZirnbauer97}

Developing an effective (response) theory of (topological) superconductors 
is beset with difficulties, at the utmost, connected to the chargeless nature of
their low-energy quasi-particles. A known approach based on topological field theories 
is to introduce new Majorana fields for these low energy degrees of freedom. 
\cite{Hansson} Other options involve cleverly constructing the topological superconducting phase
by dimensional reduction \cite{PhysRevB.87.134519} or using gravitational fields to infer about thermal 
\cite{PhysRevB.85.045104} and viscous\cite{PhysRevLett.107.075502,PhysRevD.88.025040,PhysRevB.84.085316}
responses. The gravitational approach, which is the main focus of this manuscript, has a strong
appeal due to recent advances in relating the geometrical and entanglement
properties of these systems. \cite{PhysRevB.88.195412,PhysRevB.89.195144}
The latter, in particular, has been shown to encode subtle topological characteristics of these phases. 
\cite{Zaletel}

The paper is organized as follows. In Sec. \ref{sec:Augum} we introduce the physics of defects in superconductors with broken time-reversal symmetry.
We explain how to calculate topological invariants from the single-particle Hamiltonian and relate the (3+1)D point defect case with a higher dimensional (4+1)D
situation with a line-defect, the latter case being the starting point to define our effective field theory.
In Sec. \ref{sec:EffectiveFT}, our main section, we concretelly describe our gravitational action. We show how frame-rotation invariance of our action leads to
the introduction of chiral modes living along the line defect which, upon dimensional reduction, back to (3+1)D, describe the hedgehog-bound MBS. Finally in 
Sec. \ref{sec:conclusion} we present our concluding remarks.

\section{Topological defects and dimensional augumentation \label{sec:Augum}}

\subsection{Bogoliubov de Gennes Hamiltonian and dimensional argumentation}

\begin{figure}[t!]
\begin{centering}
\includegraphics[width=0.4\textwidth]{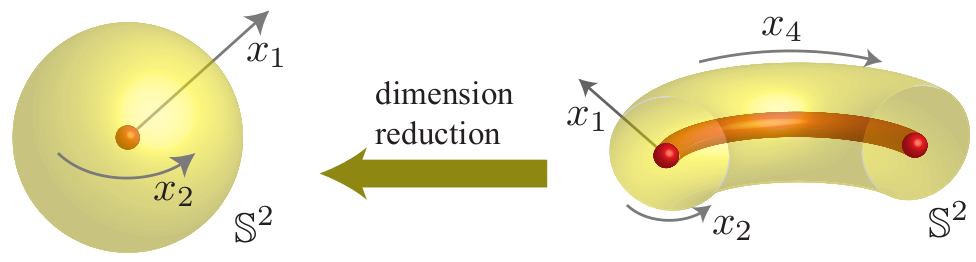}

\par\end{centering}
\protect\caption{Dimensionally reducing a line defect in (4+1) dimensions to a point defect in (3+1) dimensions. 
(Left) The point defect (red dot) is spatially surrounded by a sphere $\mathbb{S}^2$ (yellow shell).
$x_{1}$ is the radial parameter, $x_{2}$ is the azimuthal one and $x_{3}$, the
polar parameter, is not shown. (Right) The line defect (red line) parallel to the $x_4$ direction
sits inside a cylindrical three-dimensional hypersurface, 
where each cross section is a sphere $\mathbb{S}^2$ not intersecting the line defect.
}\label{fig1}
\end{figure}

We begin by reviewing the hedgehog defects discussed in Ref.\ \onlinecite{PhysRevLett.104.046401}
in terms of a non-interacting fermionic Hamiltonians. 
They are classical fields of certain order parameters, varying adiabatically in space-time,
coupled to the Bloch Hamiltonian to be studied. Such topological defects carry extensive textures around them. 
In this case, the electronic band Hamiltonian can be written as $H\left(\mathbf{k},x\right),$
where the momentum is $\mathbf{k}=\left(k^{1},\, k^{2},\, k^{3}\right)^{T}$
and the space-time coordinates $x=\left(\mathbf{x},t\right)$ characterize the defect.
Concretely, a representative for the class D Bogoliubov-de Gennes (BdG) Hamiltonian is
\begin{equation}
H\left(\mathbf{k},x\right)=\boldsymbol{\Gamma}\cdot\mathbf{k}+\boldsymbol{\Lambda}\cdot\mathbf{n}\left(x\right),\label{eq:Hamilt}
\end{equation}
where $\mathbf{n}\left(x\right)=\left(m\left(x\right),\, \mbox{Re}\Delta\left(x\right),\, \mbox{Im}\Delta\left(x\right)\right)$ 
combines the Dirac band gap $m$ with the superconducting order parameter $\Delta$, 
and the $\boldsymbol\Gamma=(\Gamma_1,\Gamma_2,\Gamma_3)$ and $\boldsymbol\Lambda=(\Lambda_1,\Lambda_2,\Lambda_3)$ 
matrices obey the Clifford algebra $\{\Gamma_i,\Gamma_j\}=\{\Lambda_i,\Lambda_j\}=2\delta_{ij}$ and $\{\Gamma_i,\Lambda_j\}=0$.
In the case of point defects in (3+1)D, we have a $d=3$-dimensional
Brillouin zone $BZ^3$ and defines a $D=d-1=2$ dimensional (spherical)
surface $\mathbb{S}^2$ around a point defect (see figure~\ref{fig1}.) This leads to a $\mathbb{Z}_{2}$ topological
classification, according to the periodic table of defects,\cite{PhysRevB.82.115120}
signaling the presence or absence of protected Majorana bound states. 
The appearance of the MBS is guaranteed by the non-trivial bulk invariant $(-1)^{\nu}$, where
\begin{align}\nu=\frac{2}{3!}\left(\frac{i}{2\pi}\right)^3\int_{\mathbb{S}^2}\int_{BZ^3}Q_5\in\mathbb{Z}_2\end{align} 
and $Q_5=\mbox{Tr}\left[\mathcal{A}d\mathcal{A}+(3/2)\mathcal{A}^3d\mathcal{A}+(3/5)\mathcal{A}^5\right]$
is the Chern-Simons 5-form and $\mathcal{A}_{mn}=\langle u_m|du_n\rangle$ is the Berry connection
constructed from the occupied BdG-states $|u_m({\bf k},x)\rangle$ of $H({\bf k},x)$.

In the context of bulk topological  insulators and superconductors 
(i.e., those without topological defects), topological states characterized by a $\mathbb{Z}_2$ topological invariant 
are closely related to (in fact, ``descend from'') their higher-dimensional parent state  
characterized by a $\mathbb{Z}$ topological invariant.  
Here in our context, we also found it to be convenient to consider a ``dimensional augmentation''.  
In this case the original point defect is extended into a line in (4+1)D, which, being a (1+1)D system, 
should support a simple chiral theory as its bound state.  

We thus augment the space with a further direction, in which case the
Brillouin zone is extended to 4D. The BdG Hamiltonian
may be written as in \eqref{eq:Hamilt} but with $\mathbf{k}=\left(k^{1},\, k^{2},\, k^{3},\, k^{4}\right)^{T}$.
This dimensional ``augmentation'' is done such that the
direction of $k^{4}$ defines now a line crossing the former point-defect, which now becomes a line-defect (see Fig.\ \ref{fig1}). 
The dimension of the sphere that wraps the defect is still $D=2$ such that now the defect dimension is $\delta=d-D=4-2=2$.
According to Ref.\ \onlinecite{PhysRevB.82.115120}, such an object is classified by an integer topological invariant
\begin{align}\nu=\frac{1}{3!}\left(\frac{i}{2\pi}\right)^3\int_{\mathbb{S}^2}\int_{BZ^4}\mbox{Tr}\left(\mathcal{F}^3\right)\in\mathbb{Z}\label{3rdChernnumber}\end{align}
for $\mathcal{F}=d\mathcal{A}+\mathcal{A}^2$ is the Berry curvature, and $BZ^4$ is the 4D Brillouin zone. 
This counts the number of Majorana chiral modes along the defect. In the following, 
we will consider a field theory describes topological excitations along the defect line. 
Through a subsequent compactification procedure, we will infer the effective field theory description of the (3+1)D theory with the point defect.

\subsection{Topological defects}

Consider now a class D system in (3+1)D with a point-defect structure. 
This is described by Eq.\ (\ref{eq:Hamilt}). 
The vector $\hat{n}\left(x\right)=\mathbf{n}\left(x\right)/\left|\mathbf{n}\left(x\right)\right|$
defines a {\em hedgehog} looking vector field around the point defect. The winding of
this vector field determines the presence or absence of particles bound
to the defect. A unit winding corresponds to a quantum vortex across a superconducting
interface between a topological and trivial insulator, an object known to bind a Majorana zero-mode. 

Extending to (4+1)D, time reversal symmetric topological insulators are $\mathbb{Z}$ 
classified \cite{NewJPhysRyu,Kitaevtable08} and the 3D interface between a 4D bulk 
primitive topological insulator (with index $\pm1$) and a trivial insulator hosts a 
single Weyl node. In the presence of superconductivity, a quantum vortex line through the 3D hyper-interface binds a chiral Majorana mode.
Unlike the anti-periodic boundary condition of a fermion ring in real 3-space where 
fermions physically rotate by $2\pi$ going around a cycle enclosing the vortex line, 
compactifying the fourth dimension simply closes the vortex loop with a periodic boundary condition on its 
chiral Majorana mode. The zero-energy zero-momentum Majorana mode is associated with the 
protected Majorana bound state at the point-defect in 
(3+1)D, surviving as the lowest energy mode after compactification.

We introduce a differential 2-form $\theta=\frac{1}{2}\theta_{\mu \nu}dx^{\mu}\wedge dx^{\nu}$
with $\theta_{\mu \nu}=\frac{1}{2} \hat{n} \cdot \partial_{\mu}\hat{n}\times\partial_{\nu}\hat{n}$. Here, $\mu=0,...,4$
in the space-time coordinates. Again, its winding counts the number
of zero-modes along the line defect and the factor of $1/2$ was introduced
such that this count matches the chiral central charge of the modes along 
the line as
\begin{align}
c_-
\equiv
c-\bar{c} &=  \frac{1}{4\pi}\int_{\mathbb{S}^{2}} \theta
\label{eqn:centralcharge 1}
\end{align}
where $c,\bar{c}$ are the central charges of the left and right moving modes, and $\mathbb{S}^2$ 
is the spherical surface that surrounds the vortex line (see Fig.\ \ref{fig1}). The chiral central charge 
relates to thermal transport behavior~\cite{PhysRevB.55.15832, NuclPhysB.636.568} by equating the energy current at temperature $T$ to $I=(\pi/12)c_-T^2$.

Equation \eqref{eqn:centralcharge 1} may be related to the bulk's Berry connection, which also
contains the information on the quasi-particles living at the defect.
\cite{PhysRevLett.104.046401}

The chiral central charge of the gapless Majorana mode along the vortex line relates to the winding of the differential 
1-form as well as the integral invariant \eqref{3rdChernnumber} by the topological index theorem
\begin{align}
\frac{1}{4\pi}\int_{\mathbb{S}^{2}}\theta
 = \frac{1}{2}\frac{1}{3!}\left(\frac{i}{2\pi}\right)^{3}\int_{\mathbb{S}^{2}}\int_{BZ^{4}}\mbox{Tr}\left[\mathcal{F}^{3}\right],
\label{eqn:centralcharge 2}
\end{align}

\begin{figure}[htbp]
\centering\includegraphics[width=0.2\textwidth]{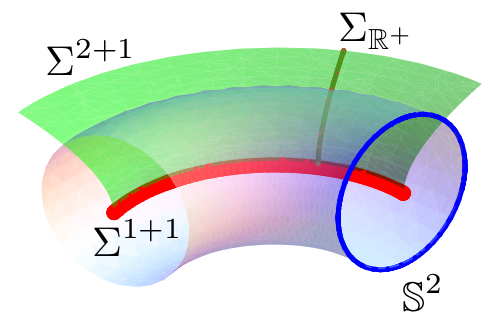}
\caption{Foliation of the (4+1)D space into $\mathbb{S}^2\times\Sigma^{2+1}$. The spherical surface $\mathbb{S}^2$ (blue) surrounds the vortex line (red). The surface $\Sigma^{2+1}$ (green) orthogonally intersects $\mathbb{S}^2$ at a point and terminates along the vortex line.}\label{fig2}
\end{figure}

The (4+1)D space can be foliated into $\Sigma^{4+1}=\mathbb{S}^2\times\Sigma^{2+1}$ 
(see Fig.\ \ref{fig2}). $\mathbb{S}^2$ is the spherical surface that encloses the point defect in (3+1)D and wraps the vortex line in (4+1)D.
$\Sigma^{2+1}$ is an open semi-infinite surface that terminates along the line defect, orthogonal to $\mathbb{S}^{2}$ at every point.
It may be decomposed as $\Sigma^{2+1}=\Sigma_{\mathbb{R}^{+}}\times\Sigma^{1+1}$.
Here $\Sigma^{1+1}$ encompasses the time ($x^{0}$) and line-defect
($x^{4}$) directions where our (1+1)D chiral Majorana modes live.
This defines a (1+1)D conformal field theory (CFT). 
$\Sigma_{\mathbb{R}^{+}}$
is the positive radial direction orthogonal to the defect sphere and
ends at the sphere's origin. In summary one may write
\begin{equation}
\Sigma^{4+1}=S^{2}\times\Sigma_{\mathbb{R}^{+}}\times\Sigma^{1+1}.
\end{equation}

\section{Effective gravitational theory \label{sec:EffectiveFT}}

\subsection{Coupling between defects and gravity}
At this point, we are ready to introduce an ansatz for the gravitational
response of this system. We fix the ansatz by a twofold argument.
The first requirement is that the action is topological in nature. 
This implies that the (4+1)D action involves a differential 5-form as its integrand which is independent of the metric (or the volume form).
The second is that the action should reflect the chiral central charge along the line defect
as this is a direct consequence of the topological field theory.
The following is the unique gravitational action that obeys these requirements in the presence of
the static order parameter field $\theta$,
\begin{align}
 \label{eq:Action}
S & =  \frac{1}{4\pi}\int_{\mathbb{S}^{2}}\theta\int_{\Sigma^{2+1}}Q_{3}^{\omega}
 \\
 & = \frac{1}{4\pi}\left(c-\bar{c}\right)\int_{\Sigma^{2+1}}Q_{3}^{\omega}, 
 \nonumber 
\end{align}
where  $Q_3^\omega$ is the gravitational Chern-Simons 3-form
such that $dQ_{3}^{\omega}=\mbox{Tr}\left[\mathcal{R}\wedge\mathcal{R}\right]$ is the 
second Chern form, with $\mathcal{R}$ the Riemann curvature tensor.

In order to exploit the similarities between tangent bundles and internal
bundles, we choose to parametrize the manifold in non-coordinate basis
in terms of local frame fields (also known as \textit{vielbeins})
$e^{a}$. These are vector valued one-forms whose vector components
run along the local bases as $a=0,...,4$ and which satisfy the local
orthogonality condition on the manifold $e_{\mu}^{a}e_{\nu}^{b}g^{\mu\nu}=\eta^{ab}$,
where $g$ is the manifold metric and $\eta$ is the local flat (Minkowski) metric. 
In terms of the local basis, the main geometric quantity of our interest,
the Riemann curvature, is written as a tensor-valued two form as
\begin{eqnarray}
\mathcal{R}_{\ b}^{a} & = & d\omega_{\ b}^{a}+\omega_{\ c}^{a}\wedge\omega_{\ b}^{c}
\end{eqnarray}
where $\omega=\omega_\mu dx^\mu$ is the spin connection, generated by the affine (Christoffel, in coordinate basis) connection
\begin{equation}
\omega_{\mu}^{ab}=e_{\nu}^{a}\partial_{\mu}e^{\nu b}+e_{\nu}^{a}e^{\rho b}\Gamma_{\rho\mu}^{\nu}
\end{equation}
for $\Gamma_{\rho\mu}^\nu=g^{\nu\kappa}\left(\partial_{\rho}g_{\mu\kappa}+\partial_{\mu}g_{\rho\kappa}-\partial_{\kappa}g_{\rho\mu}\right)/2$.
It arises as the connection which corrects derivatives and parallel transports of frame
fields in the manifold. It is intimately related to spin angular momentum
and also allows for the covariant differentiation of spinors. In terms
of the spin-connection 1-form $\omega$, the Riemann
curvature is similar to a non-Abelian gauge field strength $\mathcal{R}=d\omega+\omega\wedge\omega$
and $Q_{3}^{\omega}=\mbox{Tr}\left[\omega\wedge\left(d\omega+\frac{2}{3}\omega\wedge\omega\right)\right]$
is just the usual Chern-Simons form. 

Topological field theories with $SU(N)$ gauge groups which have some similarities with Eq.\ (\ref{eq:Action})
were previously studied, e.g.,  in Refs.\ \onlinecite{Jouko, Nair:1991ab}.
In the following, we will derive, from the action (\ref{eq:Action}), the properties of defects,
which we show are consistent with the localized Majorana zero modes at the defects.

\subsection{Gauge Invariance}

We aim at describing the physics of the modes living at the line defect.
From our previous discussions, the line defect acts as the edge of the
open manifold and consists of the radial, time and defect directions
(which we call $x^{1},\, x^{0}$ and $x^{4}$ for concreteness.) 
It is a known fact that the Chern-Simons action is not gauge invariant
in open manifolds. The restoration of gauge invariance demands the
introduction of extra chiral fields living at the edge of the manifold
\cite{WenEdge} giving rise to the boundary physics. 

So we start addressing the gauge transformation properties of our action.
Since the spin-connection is defined in a coordinate independent basis,
$Q_{3}^{\omega}$ is reparametrization invariant. It is not invariant,
on the other hand, under Lorentz transformations of the frame fields\cite{PhysRevB.85.184503}.
This rotation of the frame fields in our system works as an $SO(4,1)$
gauge transformations of the spin-connection
\begin{equation}
\omega\rightarrow O^{-1}\omega O+O^{-1}dO.\label{eq:gtransf}
\end{equation}

Under such a transformation we have\cite{PhysRevB.85.184503}
\begin{eqnarray}
S & \rightarrow & S+\delta S\nonumber \\
\delta S & = &
-\frac{1}{12\pi}\int_{\Sigma^{4+1}}\theta\wedge
\mbox{Tr}\left[\left(dO\right)O^{-1}\right]^{3}
\nonumber \\
 &  & 
 -\frac{1}{4\pi}\int_{\partial\Sigma^{4+1}}\theta\wedge\alpha_{2},
\end{eqnarray}
where $\alpha_{2}=\mbox{Tr}\left[ \left(dO\right)O^{-1}\wedge\omega\right] $.
To maintain gauge invariance, the factor $\delta S$ has to be compensated by the introduction of
extra degrees of freedom.

The general form of $\delta S$ dictates which type of degrees
of freedom needs to be introduced. A detailed study of the most
general behavior of $\delta S$ is thus imperative. We start by foliating the (4+1)D
manifold in spheres around the point defect. Integrating over the spherical surface around the
defect gives
\begin{align}
\delta S & = -\frac{c_-}{12\pi}\int_{\Sigma^{2+1}}\mbox{Tr}\left[\left(dO\right)O^{-1}\right]^{3} -\frac{c_-}{4\pi}\int_{\Sigma^{1+1}}\alpha_{2},\label{action1}
\end{align}
where $c_-=c-\bar{c}$ for the chiral central charge seen in \eqref{eqn:centralcharge 1}.

The first term corresponds to a Wess-Zumino-Witten (WZW) action
in an open 3-manifold while the last term couples the spin-connection
to the current $\left(dO\right)O^{-1}$ along the line-defect. 
The topological nature of the WZW action allows us to restrict our consideration to the $SO(4)$ subgroup in $SO(4,1)$, which deformation retract to $SO(4)$ 
since the boost directions are contractible. As a consequence, we assume without loss of generality that all transformations $O$ are $SO(4)$ rotations.

Rotations in four dimensional manifolds may be decomposed as rotations
of pairs of planes. In particular, a general $SO(4)$ rotation may
be separated into the product of, up to a pair of global inversions,so-called
left- and right-isoclinic rotations. \begin{align}O=AB=(-A)(-B)\end{align} These correspond to rotations where 
both pairs of planes rotate by the same angle (for $A$) or opposite angles (for $B$).
These isoclinic rotations are equivalent to unit quaternion elements, which themselves
are elements of the $SU(2)$, denoted by lower case letters $a$ and $b$ respectively. This gives the well known double cover,
\begin{equation}
SO(4)\equiv\frac{SU(2)\times SU(2)}{\mathbb{Z}_{2}};\label{eq:doubcov}
\end{equation}
where $\mathbb{Z}_2$ is the group of inversions generated by $(-1,-1)$ in $SU(2)\times SU(2)$.
Thus, we may separate the $SO(4)$ WZW action in a pair of $SU(2)$ WZW ones as
\begin{align}
  &  \mbox{Tr}_{SO(4)}\left[\left(dO\right)O^{-1}\right]^{3}
  \nonumber \\
 =& \mbox{Tr}_{SO(4)}\left[\left(dA\right)A^{-1}\right]^{3}
 +\mbox{Tr}_{SO(4)}\left[\left(dB\right)B^{-1}\right]^{3} \\
 = & 2\left\{
 \mbox{Tr}_{SU(2)}\left[\left(da\right)a^{-1}\right]^{3}
 +\mbox{Tr}_{SU(2)}\left[\left(db\right)b^{-1}\right]^{3}
 \right\}
 \nonumber,
\end{align}
where we noticed that the $SO(4)$ trace is twice of that of a $SU(2)$ one. For completeness and to address the unfamiliarized reader,
we present in the appendix a detailed description of this mapping

A further caveat must be taken into account. The compactification
of the line defect shrinks one of the rotation
directions and reduces the $SO(4)$ group to $SO(3)=SU(2)/\mathbb{Z}_2$.
It contains the rotations in the dimension reduced 3+1D physical space.
This is just the diagonal subgroup of $SO(4)$ in {Eq.\ \eqref{eq:doubcov}, i.e. $a=b$. This means that
the compactification of the defect line confines the two $SU(2)$ theories  into a single one. 
Thus the first term of the action \eqref{action1} becomes
\begin{align}
  &-\frac{4c_-}{12\pi}\int_{\Sigma^{2+1}}\mbox{Tr}_{SU(2)}\left[\left(da\right)a^{-1}\right]^{3}\nonumber\\
 =& -\frac{2}{12\pi}\int_{\Sigma^{2+1}}\mbox{Tr}_{SU(2)}\left[\left(da\right)a^{-1}\right]^{3},
\end{align}
where we identified $c_-=1/2$ for a single chiral Majorana fermion at the line defect.
The $SO(4)$ WZW theory is now reduced to a single $SU(2)_2$,
and the overall factor of 2 fixes the level of the affine Kac-Moody 
current running along the boundary of $\Sigma^{2+1}$, which is the (1+1)D vortex line.

Finally we see,
\begin{eqnarray}
\delta S & = & -\frac{2}{12\pi}\int_{\Sigma^{2+1}}\mbox{Tr}\left[\left(da\right)a^{-1}\right]^{3}\nonumber \\
 &  & -\frac{2}{4\pi}\int_{\Sigma^{1+1}}\mbox{Tr}\left[\left(da\right)a^{-1}\wedge\omega\right],
\end{eqnarray}
where the trace is understood to be take over $SU(2)$ matrices. The original action must be modified
to compensate for this. We address this issue in the next section, defining the edge theory.

\subsection{Edge Theory}

As discussed, the line-defect acts as a boundary to the manifold, rendering the action defined in Eq.\ (\ref{eq:Action}) not gauge invariant. 
Typically one fixes then a gauge and, solving the equations of motion for the resulting constraint, obtains the action for the edge theory.
One says that the loss of gauge invariance sets free extra degrees of freedom, which then are allowed to become dynamical.
\cite{WenEdge} 

In the present context, gauge invariance amounts to invariance under Lorentz transformations.
This invariance is closely connected to energy-momentum conservation. We would like to be able to 
do Lorentz transformations at the boundaries as well as in the bulk and gauge fixing is therefore not desirable.

We simply notice that introducing the proper set of degrees of freedom 
we may recover the gauge invariance desired. 
This is achieved, in our case, by substituting the original
action (\ref{eq:Action}) by 
\begin{align}
S = & \frac{1}{4\pi}\int_{\Sigma^{4+1}}\theta\wedge\mbox{Tr}\left(Q_{3}^{\omega}\right)\nonumber \\
   & +\frac{4}{12\pi}\int_{\Sigma^{4+1}}\theta\wedge\mbox{Tr}\left(J^{3}\right)\nonumber\\&+\frac{4}{4\pi}\int_{\partial\Sigma^{4+1}}\theta\wedge\mbox{Tr}\left(J\wedge\omega\right) ,
\end{align}
where 
\begin{equation}
J=\left(ds\right)s^{-1},
\quad
s \in SU(2)
\end{equation}
is the $SU(2)$ current operator of the edge theory for $s\in SU(2)$ along the vortex line $\Sigma^{1+1}$. Under
a gauge transformation by an $SO(3)$ rotation of the frame-field, the
gravitational part transforms as follows from Eq.\ (\ref{eq:gtransf}) while
$s\rightarrow as$. The changes in the two counterterms cancel the
changes in the first one, giving the desired invariance. 

The boundary condition (as specified by the sign of the winding number) of the defect field $\theta$ 
determines the chirality of the current operators. 
This theory still lacks dynamics, as all defects in topological phases, in the form
of a non-linear sigma model. In the presence of a  kinetic term, however the sign of the WZW action (the sign in front of the level) 
fixes the conservation equations for the corresponding chirality.

We are not going to consider further the dynamics along the line defect $\Sigma^{1+1}$, 
which is treated to be static in this manuscript. 
After integrating with static defect field $\theta$ over the spherical surface around it, 
we have an $SU(2)_{2}$ WZW term on the (2+1)D space $\Sigma^{2+1}$ orthogonal to the surface around the line defect 
together with a coupling between the bulk geometry and the edge modes. 
Explicitly, we obtain
\begin{align}
S  = \frac{2}{12\pi}\int_{\Sigma^{2+1}}\mbox{Tr}\left(J^{3}\right)+\frac{2}{4\pi}\int_{\partial\Sigma^{2+1}}\mbox{Tr}\left(J\wedge\omega\right) 
\label{eq:EdgeS}
\end{align}
where the integration of $\theta$
over $\mathbb{S}^{2}$ gives the chiral central charge $c_-=1/2$ and is absorbed in the coefficient.

The last term in \eqref{eq:EdgeS} acts as
a coupling between the $SU(2)_2$ current $J$ and the bulk geometry through the spin connection
$\omega$. The vortex line $\Sigma^{1+1}$ is parametrized by $x^4$ and $x^0=t$ (see figure~\ref{fig1}). 
With these coordinates, the coupling becomes
\begin{align}
 \frac{2}{4\pi}\int_{\Sigma^{1+1}}\mbox{Tr}\left(J\wedge\omega\right) 
 = \frac{1}{2\pi}\int dx^{4}dt\,
 \mbox{Tr}\left(J_{0}\omega_{4}-J_{4}\omega_{0}\right) .
\label{Jomegacoupling}
\end{align}
In a almost-flat geometry, the frame field components are
\begin{align}
e_{\ \mu}^{a} \approx \delta_{\ \mu}^{a}+\frac{1}{2}h_{\ \mu}^{a}
\end{align}
and, for symmetric perturbations, the spin-connection reduces to
\begin{equation}
\omega_{\mu b}^{a}=
\frac{1}{2}\eta^{a\rho}\left(\partial_{b}h_{\rho\mu}-\partial_{\rho}h_{\mu b}\right).
\end{equation}
After compactification, $x^{4}=x^{4}+2\pi R$, for $ R\to0$,
and all $x^{4}$ derivatives should vanish and the metric to decouple
from the other directions, such that $\omega_{4b}^{a}=0$. 
Then the coupling \eqref{Jomegacoupling} simplifies into
\begin{equation}
-\frac{1}{2\pi}\int dx^{4}dt\,
\mbox{Tr}\left(J_{4}\omega_{0}\right).
\end{equation}

The dynamical spin-density then reads
\begin{equation}
l^{b}_{0a} \equiv
\frac{2}{\sqrt{g}}\frac{\delta S}{\delta\omega_{0b}^{a}}
=-\frac{1}{\pi}J_{4a}^{b}+\ldots\label{eq:spindens}
\end{equation}
where the unspecified terms are bulk contributions and are suppressed by its gap.

\begin{figure}[htbp]
\centering\includegraphics[width=0.3\textwidth]{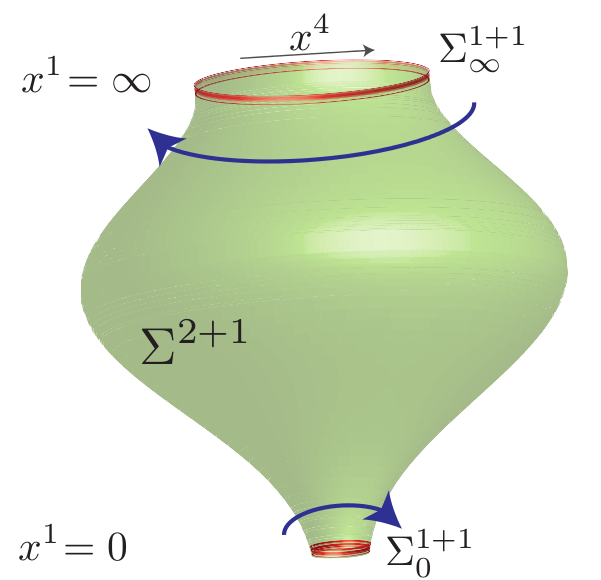}
\caption{Non-locally correlated angular momentum between zero energy Majorana bound states (red rings) at $0$ and $\infty$ 
through the $SU(2)$ field $s$ in the bulk.}\label{fig3}
\end{figure}

This angular momentum density may be rewritten in terms of the $SU(2)$ fields (we omit the matrix indices)

\begin{equation}
l_0=-\frac{1}{\pi}J_{4}=-\frac{1}{\pi}\partial_{4}\log s.
\end{equation}
We conjecture that this expression represents a non-local storage
of angular momentum between two defects, one placed at the origin
and the other at infinity (see Fig.\ \ref{fig3}). 
The defect at infinity is not seen because
in Eq.\ (\ref{eq:EdgeS}) as only the origin boundary at $x^1=0$ was taken into account.
The more general expression would include the other Majorana mode
\begin{equation}
l_0\sim\partial_{4}\log s|_{\infty}-\partial_{4}\log s|_{0},
\end{equation}
which reduces to the previous result if $s|_{\infty}$ is a constant.
The total angular momentum $L$ is obtained by integrating over $x^{4}$,
and we may write it suggestively as
\begin{equation}
L=\frac{1}{\pi}\int_{0}^{2\pi R}dx_{4}\int_{0}^{\infty}dx_{1}\partial_{1}\partial_{4}\log s.
\end{equation}

The dimensional augumentation resulted in a ``fattening'' of the string connecting the two defects
allowing us to follow the the angular momentum stored in the system. The semi-extensive nature of
the point defects through the bulk in these systems (which was crucial to the identification of their non-Abelian properties
in Ref.\ \onlinecite{PhysRevLett.104.046401} is implied and encoded by the bulk $SU(2)$ texture field $s$.
This is similar to a bulk-boundary correspondence, where spatially separated low energy degrees
of freedom connect to each other through the bulk high energy modes. 

The total angular momentum in a closed system is conserved. This means twisting the Majorana mode 
at one end will generate an anti-twist at the other end. We attribute this non-local angular 
momentum transfer to the extensive bulk $SU(2)$ texture. This angular momentum pump could be a 
gravitational version of the Thouless charge pump across an insulating chain~\cite{Thouless,NiuThouless} 
and the fermion parity pump across a topological $p$-wave superconducting chain.
\cite{Kitaevchain,PhysRevB.82.115120}

\section{Conclusion \label{sec:conclusion}}

We have studied point defects in class D topological superconductors
in (3+1) space-time dimensions from the point of view of topological
field theories. From symmetry arguments, we have proposed a minimal gravitational Chern-Simons
model coupled with a bulk texture in the extended (4+1)D space.

We have shown that the point defect extends to a line
defect after the ''dimensional augumentation'', and acts as an effective boundary or vortex line.
In order to recover Lorentz invariance, one is then forced to introduce extra boundary degrees of freedom.
This is given by a chiral $SU(2)_2$ WZW theory coupled to the bulk geometry through the spin connection.
The chirality of the new fields is fixed by the bulk texture. Under periodic boundary condition, the compactified 
vortex line traps a single zero energy Majorana mode, and through bulk coupling, we have shown non-local angular 
momentum correlation between spatially separated defects.

As a final concluding remark, we consider the possibility of higher winding of
the defect field. In this case, integrating over this field yields an $SU(2)$
WZW theory at level $2n$, for a $n$-fold winding, 
since the coefficient of the WZW action scales linearly with the winding number. 
On the other hand, one would expect the vortex line to consist of
$n$ copies of $SU(2)_2$ theories and hold $n$ chiral Majorana modes. We suspect this could be recovered 
by first relating $SU(2)_{2n}$ and $SU(2n)_2$ through the level rank duality, then regard  $(SU(2)_2)^n$ as a conformal embedding in 
$SU(2n)_2$. In this case, the theory captures only the spin part while other flavor degrees of freedom are either uncoupled
to gravity or are confined by the locality of electrons.

\section*{Acknowledgements}

P.L.S.L. acknowledges support from FAPESP under grants 2009/18336-0
and 2012/03210-3.
S. R. is supported by Alfred P. Sloan Foundation.

\appendix*
\section{The $SO(4)$ double cover}
Here we present a more detailed discussion of the identification $SO(4)\equiv\frac{SU(2)\times SU(2)}{\mathbb{Z}_{2}}$. In particular
we present an explicit formula mapping the $SO(4)$ $A$ and $B$ fields to the $SU(2)$ $a$ and $b$ ones.
Let us start thinking of rotations in a four-dimensional space. If we write a general point as $\mathbf{r}=\left(w,\, x,\, y,\, z\right)$, we may
decompose a general rotation in rotations of orthogonal planes, say $\left(w,\, x,\, 0,\, 0\right)$ and $\left(0,\, 0,\, y,\, z\right)$ for simplicity. These rotations
leave the normal vectors of the planes fixed. If these rotations turn the planes around their normals for the same angular displacement they are called
isoclinic rotations. For our particular example they may be written, for an angle $\theta$,
\begin{equation}
 O_{iso}=\left(\begin{array}{cccc}
\cos\theta & -\sin\theta\\
\sin\theta & \cos\theta\\
 &  & \cos\theta & -\sin\theta\\
 &  & \sin\theta & \cos\theta
\end{array}\right)
\end{equation}
If one exchanges the ordering of the basis vectors $y$ and $z$, we arrive at another, equally reasonable, possibility, namely
\begin{equation}
 O_{iso}^{'}=\left(\begin{array}{cccc}
\cos\theta & -\sin\theta\\
\sin\theta & \cos\theta\\
 &  & \cos\theta & \sin\theta\\
 &  & -\sin\theta & \cos\theta
\end{array}\right).
\end{equation}
Rotations with like-signs $\left(\theta,\,\theta\right)$ are called left-isoclinic while those of opposite signs $\left(\theta,\,-\theta\right)$ are called right-isoclinic. From 
the shape of the matrices in this particular case, the $SU(2)$ nature of the isoclinic rotations already becomes apparent. One may be more general, calling $A$ and $B$ the
left- and right-isoclinic matrices, they may be written
\begin{eqnarray}
 & A=\alpha_{1}-i\sigma_{y}\alpha_{2}-i\tau_{y}\sigma_{z}\beta_{1}-i\tau_{y}\sigma_{x}\beta_{2},
 \nonumber \\
 & B=\gamma_{1}-i\tau_{z}\sigma_{y}\gamma_{2}-i\tau_{y}\delta_{1}-i\tau_{x}\sigma_{y}\delta_{2} , 
\end{eqnarray}
where $\sigma$ and $\tau$ are Pauli matrices and we omitted the identity matrices.
The easiest way to see that indeed one may decompose a general rotation in four dimensions in such rotations, namely $O=AB$, is to complexify the coordinates. In this case, we
have
\begin{equation}
 \mathbf{r}=\left(x_{1},\, x_{2},\, x_{3},\, x_{4}\right)\rightarrow X=\left(\begin{array}{cc}
y & z\\
-\bar{z} & \bar{y}
\end{array}\right),
\end{equation}
where $y=x_{1}+ix_{2}$ and $z=x_{3}+ix_{4}$ and the bars represent complex conjugation. Now the most general transformation on the matrix $X$ which preserves its determinant (and as such, the modulus of $\mathbf{r}$) reads
\begin{equation}
 X^{'}=aXb,
\end{equation}
where $a$ and $b$ are $SU(2)$ transformations. Naturally, multiplying both $a$ and $b$ by $-1$ gives the same result, which accounts for the double-cover. Now simply taking general $a$ and $b$
matrices and separately equating each one of them to the identity and expanding, one may relate these to the $SO(4)$ isoclinic rotations $A$ and $B$. At the end of a short calculation, we
find an explicit relation between them, which may be used in the fields of the main text, as follows
\begin{eqnarray}
A & \rightarrow & a=\left(\begin{array}{cc}
\alpha & \beta\\
-\bar{\beta} & \bar{\alpha}
\end{array}\right),
\\
B & \rightarrow & b=\left(\begin{array}{cc}
\gamma & \delta\\
-\bar{\delta} & \bar{\gamma}
\end{array}\right)
\end{eqnarray}
where $\alpha=\alpha_{1}+i\alpha_{2}$ (with $\alpha_{1,2}$ as defined above for $A$) and the notation follows similarly for $\beta$, $\delta$ and $\gamma$.

\bibliographystyle{apsrev4-1}

\end{document}